\documentclass[11pt,twoside]{article}
\usepackage{asp2010}

\resetcounters

\begin{document}

\title{Requirements for the Analysis of Quiet Sun Internetwork Magnetic Elements with EST and ATST}
\author{D. Orozco Su\'arez,$^1$ L.R. Bellot Rubio,$^2$ and Y. Katsukawa$^1$
\affil{$^1$National Astronomical Observatory of Japan, 2-21-1 Osawa, Mitaka, Tokyo 181-8588, Japan}
\affil{$^2$Instituto de Astrof\'{\i}sica de Andaluc\'{\i}a (CSIC), Apdo.\ de Correos 3004, 18080 Granada, Spain}}

\begin{abstract}
The quiet Sun internetwork is permeated by weak and highly inclined
magnetic fields whose physical properties, dynamics, and magnetic
interactions are not fully understood. High spatial resolution
magnetograms show them as discrete magnetic elements that
appear/emerge and disappear/cancel continuously over the quiet Sun
surface. The 4-m European Solar Telescope (EST) and the Advanced
Technology Solar Telescope (ATST) will obtain two-dimensional, high
cadence, high precision polarimetric measurements at the diffraction
limit (30 km). Here, we compile the basic requirements for the
observation of internetwork fields with EST and ATST (field of view,
cadence, instrument configuration, etc). More specifically, we
concentrate on the field-of-view requirements. To set them we analyze
the proper motion of internetwork magnetic elements across the solar
surface. We use 13 hours of magnetograms taken with the Hinode
satellite to identify and track thousands of internetwork magnetic
element in an isolated supergranular cell. We calculate the velocity
components of each element and the mean distance they travel. The
results show that, on average, magnetic elements in the interior of
supergranular cells move toward the network. The radial velocity is
observed to depend on the distance to the center of the supergranule.
Internetwork magnetic elements travel 4\arcsec\/ on average. These
results suggest that ATST and EST should cover, at least, one
supergranular cell to obtain a complete picture of the quiet Sun
internetwork.
\end{abstract}

\section{Introduction: Properties of Quiet Sun Internetwork Magnetic Fields}

Obtaining a clear picture of the solar magnetism and of the various
physical processes taking place in the solar atmosphere strongly
relies on the precise determination of the magnetic field vector. This
is very true for the tiny magnetic structures that permeate the
quieter areas of the solar photosphere, the so-called internetwork
fields (IN; \citealt{1975BAAS....7..346L,2008ApJ...672.1237L}). Many of the physical
properties of the IN fields have been inferred from the analysis of
steady four-dimensional images (x, y, $\lambda$, p)\footnote{Two
spatial dimensions, wavelength, and four polarization states (Stokes
I, Q, U, and V).} constructed from spectropolarimetric
measurements. For instance, IN observations taken with the
spectropolarimeter (SP; \citealt{2001ASPC..236...33L}) of the 50-cm
telescope onboard the Hinode satellite \citep{2007SoPh..243....3K}
show that IN magnetic fields are weak and have large inclinations to
the local vertical (e.g., \citealt{2012arXiv1203.1440O}). However, even at the best spatial resolution
achieved so far in the IN (0\farcs15 with \textsc{Sunrise}/IMaX,
\citealt{2011SoPh..268....1B,2011SoPh..268...57M}), the amplitude 
of the polarization signals are very small and thus exposed to harmful
noise effects. Moreover, the magnetic structures are still partially
unresolved. Recent inversions of Stokes profiles indicate that the
magnetic fill fractions in the IN are about 20\% at 0\farcs3
resolution \citep{2007ApJ...670L..61O}. Also, observations show that
the number and size of magnetic structures increase monotonically as
the resolution improves
\citep{2008ApJ...684.1469D,2010ApJ...723L.149D,
2010ApJ...718L.171I}. On theoretical grounds, it is expected that
magnetic structures will exist up to the resistivity scales of the
solar plasma (a few meters, e.g., \citealt{2009SSRv..144..275D} 
and \citealt{2009ApJ...693.1728P}).

In addition, the magnetic fields of the IN are highly dynamic and
evolve through several physical processes such as emergence and
cancellation. However, it is difficult to study their evolution using
high-cadence raster scans from a spectropolarimeter. The reason is
that to one needs exposures of a few seconds per slit position to
achieve a polarimetric sensitivity of, say, $10^{-3} I_{\mathrm{c}}$
(or s/n~$\sim$~1000)\footnote{The signal-to-the-noise ratio (s/n) is
calculated in the continuum of Stokes I.}. Therefore, the measurements
have to be limited to very narrow regions of the solar surface, and
even so the achievable cadence is of the order of minutes. Using this
technique, it has been possible to characterize dynamic properties of
magnetic flux emergence processes (e.g.,
\citealt{2008A&amp;A...481L..33O, 2009ApJ...700.1391M}). However, 
most of the dynamic properties of the IN fields have been determined
from filter-based polarimeters which allow the observation of large
fields of view at high cadence, but with far less spectral
purity. From these measurements we know that the mean lifetime of IN
magnetic elements is about 10 minutes, although it is difficult to set
a mean lifetime for the polarization signals at high resolution (e.g.,
\citealt{2010ApJ...723L.149D}). The rms fluctuation of the horizontal 
velocity of IN magnetic elements is 1.57~km~s$^{-1}$
\citep{2008ApJ...684.1469D}. A drawback of filter-based instruments is
that the physical information extracted from the observed polarization
signals is often less accurate than that inferred from slit-based
polarimeters because the data are affected by the modest spectral
resolution and coarse wavelength sampling
\citep{2010A&A...522A.101O}.

\begin{figure}[!t]
\centering
\plotone{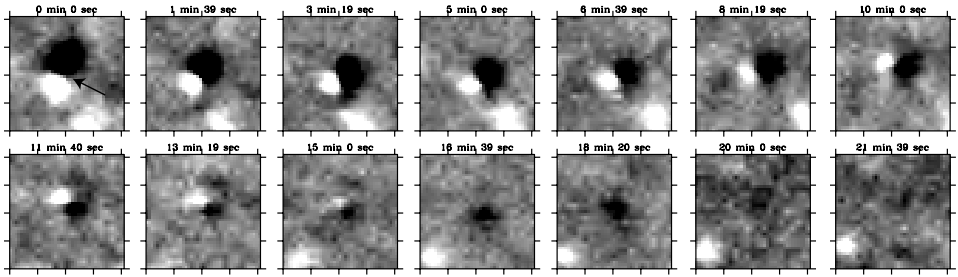}
\caption{Magnetograms recorded with Hinode's Narrowband Filter Imager 
showing the cancellation of two IN magnetic elements (left to right
and top to bottom). The total duration of the sequence is about 22
minutes. The arrow pinpoints the cancellation site or polarity
inversion line that is very small compared to the size of the
canceling magnetic elements. Note how the magnetic elements rotate
clockwise while the amplitude of the signals diminishes, suggesting
that important dynamics may take place during the cancellation. Tick
marks are separated by 1\arcsec.}
\label{fig1}
\end{figure}

Finally, the limited spatial resolution impedes the investigation of
small-scale physical processes like reconnections, jets, or
cancellations taking place in the quiet Sun IN. For instance, recent
analyses of cancellation events observed in the IN with the Hinode
spectropolarimeter show evidence of insufficient temporal and spatial
resolution to ``catch'' the important physics behind these events (see
e.g., \citealt{2010ApJ...712.1321K}). An example of a magnetic
cancellation is shown in Fig.~\ref{fig1} using \ion{Na}{I} 589.6~nm
magnetograms acquired with the Narrowband Filter Imager of Hinode at a
spatial resolution of about 0\farcs3. The cancellation takes place in
a very narrow region, right between the two magnetic elements, in the
so-called polarity inversion line. Present spectropolarimetric
measurements lack enough spatial resolution to observe cancellation
boundaries in detail. In summary, the observation of IN fields by
spectropolarimetric means is a very demanding task, but necessary to
determine the physical properties of the quiet Sun.

Large photon collecting surfaces will improve both the photometric
accuracy and the cadence of the measurements
\citep{2008LNEA....3..113C, 2010AN....331..558D}, as well as the
spatial resolution.  Thus, it will be easier to determine the magnetic
field vector from the observed polarization signals because the
changes in the Stokes profile shapes resulting from variations of the
physical quantities will be less influenced by the noise (e.g.,
\citealt{2011IAUS..273...37D}). In addition, the amplitude of the
polarization signals will be greater since the magnetic filling
factors tend to increase as the spatial resolution improves. By the
same token, however, resolution also challenges the observation of
extended areas of the solar surface at high cadence using
spectropolarimeters, because of the large number of steps in the slit
scan direction needed to cover significant fields of view. The
European Solar Telescope (EST; \citealt{2010AN....331..615C}) and the Advanced Technology
Solar Telescope (ATST; \citealt{2010AN....331..609K}) will be equipped with 4m mirrors,
thus facilitating the measurement of quiet Sun IN field. Both will
provide the photons needed to achieve high polarimetric accuracy and
temporal resolution with small exposure times\footnote{For details,
see http://www.est-east.eu and http://atst.nso.edu}.

In this contribution we aim at listing the main requirements for
observing quiet Sun internetwork fields with EST and ATST using the
current understanding about the magnetic and dynamic properties of IN
magnetic elements. We put special attention on the fields of view
needed to get a complete picture of the quiet Sun internetwork
magnetism and of the physical processes taking place there.

\section{Velocities of Internetwork Magnetic Elements and Traveled Distance}

To set up limits on the field of view (FOV) to be observed by EST and
ATST one can use the typical velocities (i.e., proper motions) and
lifetimes (or traveled distances) of IN magnetic elements as a
reference. If we are interested in analyzing individual magnetic
features from birth (emergence/submergence or appearance of flux) to
death (cancellation or disappearance), these two quantities may help
constrain the required FOV because they inform us about the mean
distance traveled by the magnetic elements in the solar surface. Most
measurements concur that IN flux concentrations show two distinct
velocity components: one of random nature, resulting from the
continual buffeting of the fields by granular convection (e.g.,
\citealt{2011A&A...531L...9M}), and a net velocity that transports
the fields from the center of supergranular cells toward their 
boundaries, i.e., toward the network.  We are interested in the
latter. However, while there exist robust estimations of the random
velocity component, about 1.57~km~s$^{-1}$, the values given in the
literature for the net velocity range from 0.2~km~s$^{-1}$
\citep{2008ApJ...684.1469D} to 0.35~km~s$^{-1}$
\citep{1987SoPh..110..101Z}. Only \cite{1998A&A...338..322Z}
investigated the spatial dependence of the velocity pattern of IN
elements and found a constant radial velocity (measured from the
center of supergranular cells outwards) of about 0.4~km~s$^{-1}$ using
low resolution (1\farcs5) and low cadence (7 min) magnetograms.

\begin{figure*}[!t]
  \centering
\epsscale{0.6}
\plotone{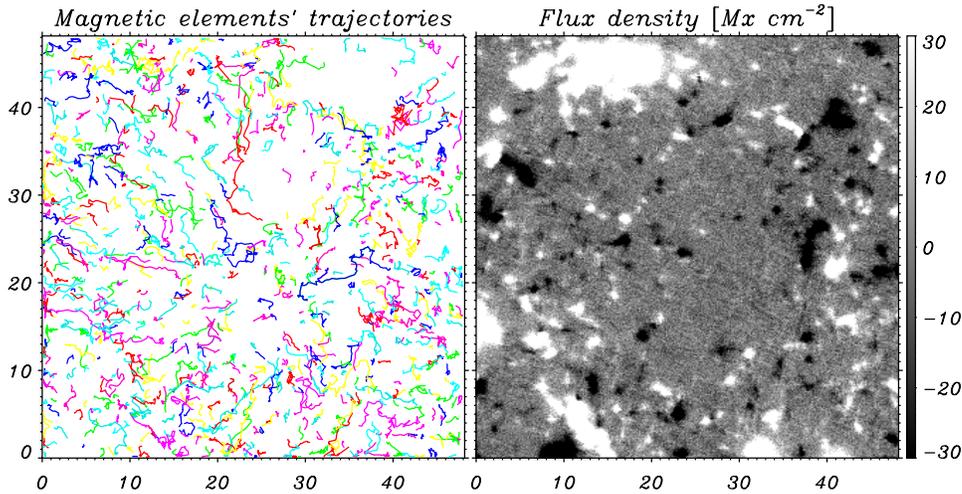}
\caption{Right: Quiet Sun magnetogram taken with Hinode's Narrowband
Filter Imager, saturated at $\pm$~30~G. The FOV covers one
supergranular cell. Left: trajectories of some of the several thousand
magnetic elements detected in 13 hours of magnetograms. The
trajectories are more linear in the inner part and more random near
the borders. Colors distinguish individual objects. Axis units are
arcsec.}
\label{fig2}
\end{figure*}

Here, we employ very high spatial resolution (0\farcs3)
\ion{Na}{I} 589.6~nm magnetograms acquired by Hinode to estimate the
net velocity component of IN magnetic elements. The magnetograms have
a cadence of 80 seconds and a FOV of approximately
93\arcsec$\times$82\arcsec, so they cover a few supergranular
cells. The noise is 6.7~G. To analyze the data we first selected a
small area of 50\arcsec$\times$50\arcsec\/ containing one
supergranular cell and the network around it. Then we detected and
tracked the proper motion of each magnetic element in consecutive
magnetograms. In total, we analyzed 13 hours of data. For detecting
and tracking magnetic elements we used the YAFTA\_10
software\footnote{YAFTA\_10 can be downloaded from
http://solarmuri.ssl.berkeley.edu/$\sim$welsch/public/software/YAFTA}
\citep[Yet Another Feature Tracking Algorithm;][]{2007ApJ...666..576D}. 
In the detection process we ignored all magnetic elements with fluxes
below 10~G and sizes smaller that 16~px$^2$.

The right panel of Figure \ref{fig2} depicts a single
50\arcsec$\times$50\arcsec\/ snapshot extracted from the magnetogram
time sequence. It contains a supergranular cell. One can clearly see
large circular polarization signals concentrated at the borders of the
image, corresponding to large flux values (i.e., network fields). In
the inner part of the image there are flux elements that show opposite
polarities and much smaller sizes than those associated with the
network. These signals correspond to internetwork magnetic fields.
The left panel shows the paths of some of the several thousand
elements detected and tracked in the magnetograms by YAFTA\_10. The
paths of the elements located in the inner part of the supergranule
are rather linear and radially aligned with respect to the center of
the image. In the network, the paths ``look'' more random and show
little linear trends. This figure alone suggests that quiet Sun
magnetic elements have two distinct velocity components, one random
and one systematic, both weighted differently depending on the
location of the element within the supergranular cell.

Once we have the positions of each magnetic element in each of the
magnetograms, we derive their horizontal velocities $\mathbf{v} =
(v_{\rm x},v_{\rm y})$. Because there is a clear radial symmetry in
the trajectory pattern we decided to translate the horizontal
velocities into the radial $v_{\rm R} = \mathbf{v}\cdot\mathbf{r} /
|\mathbf{r}|$ and transversal $v_{\rm T} =
|\mathbf{r}\times\mathbf{v}| / |\mathbf{r}|$ components with respect
to the center of the supergranular cell. Here the quantity
$\mathbf{r}$ stands for the displacement and $v_{\rm R}$ is measured
positive outwards. Using the network at a reference, the supergranule
center can be located at $(x_{\rm c}, y_{\rm c}) \sim (26\farcs6,
22\farcs1)$, i.e., very close to the center of the magnetogram
displayed in Figure~\ref{fig2}. Next we calculated the azimuthally
averaged values of the radial velocity component as a function of
distance from the center. We also calculated the mean linear distance
$\overline{|\mathbf{r}|}$ traveled by the magnetic elements detected
by the tracking algorithm.

\begin{figure}[!t]
  \centering
\plottwo{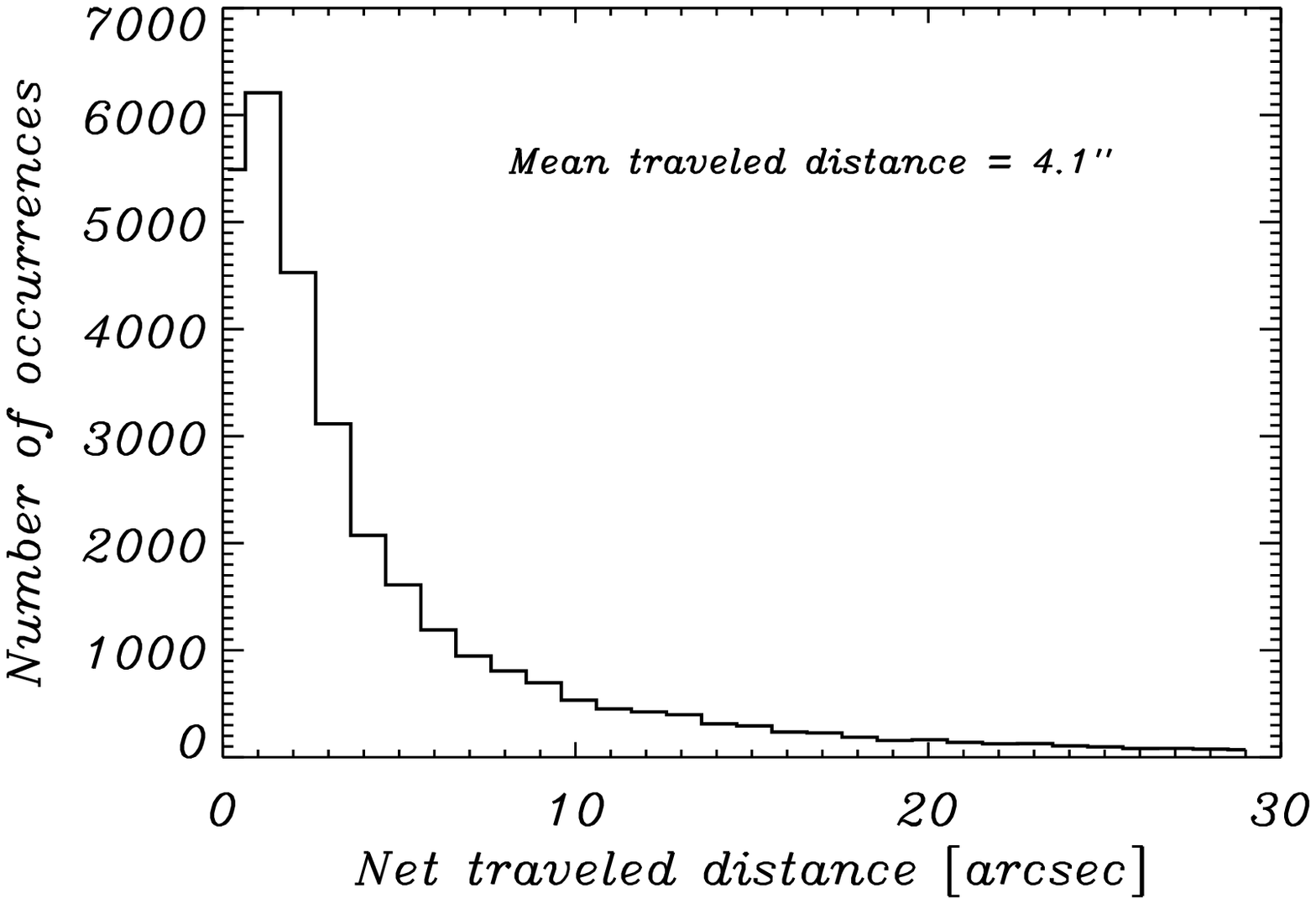}{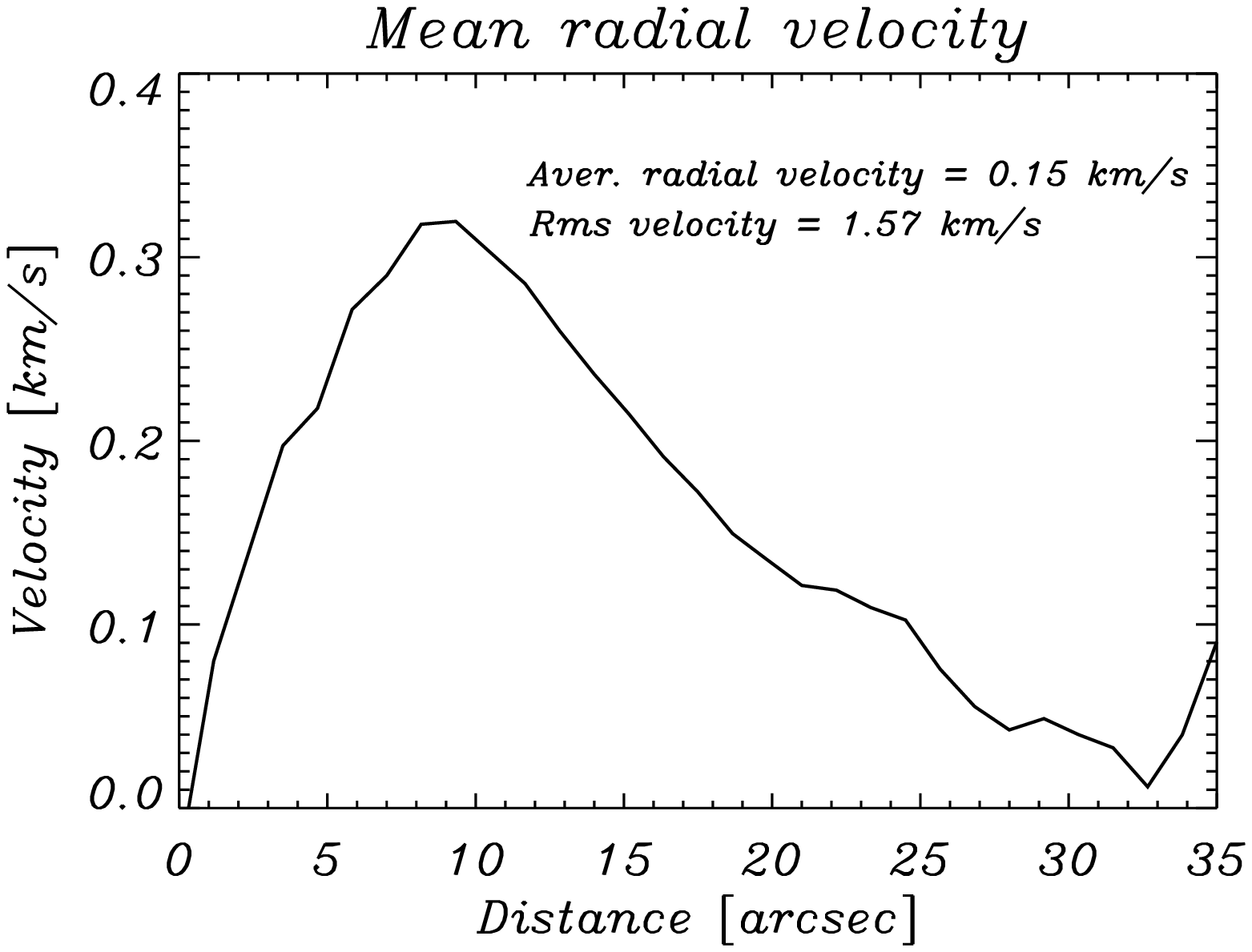}
\caption{Left: histogram of the mean linear distance traveled by 
internetwork magnetic elements. The average value is 4\farcs1.  Right:
Variation of the azimuthally averaged radial velocity component of the
magnetic elements as a function of distance (position) from the center
of the supergranule.}
\label{fig3}
\end{figure}

The results are shown in Fig.~\ref{fig3} for the average radial
velocity (right panel) and the mean traveled distance (left
panel). Interestingly, the radial component of the velocity shows a
dependence with the distance to the supergranular cell center. The
magnetic elements close to the center are almost at rest. The radial
velocity increases outwards until it reaches a maximum value of
0.32~km~s$^{-1}$ at 9\arcsec\/ from the supergranular center. Then,
it decreases monotonically. We find a mean radial velocity of
0.15~km~s~$^{-1}$, slightly smaller than previous estimations
\citep{2008ApJ...684.1469D,1987SoPh..110..101Z,1998A&A...338..322Z}. Overall,
the magnetic elements move toward the network. These results indicate
that the dynamic properties of the quiet Sun magnetic elements are
different depending on their location within the supergranular cell.

The histogram for the mean traveled distance shows a peak at small
values and has an extended tail that reaches 20\arcsec. This implies
that most magnetic elements travel small distances, although some of
them can get as far as half the diameter of a typical supergranular
cell. The mean traveled distance is 4\farcs1. It is important to keep
in mind that during the tracking process we did not take into account
the fact that small magnetic elements tend to collide with others
during their trip to the network. Such collisions give rise to
splitting of signals, merging of signals of the same polarity, and
cancellation of signals of opposite polarity. Such interactions create
``new'' elements in the sense that the signals change their properties
enough to make the tracking algorithm interpret them as new magnetic
elements. Hence, the mean distance is being underestimated and the
value given here should be considered a lower limit.

\section{Summary: Minimum Requirements to Observe IN Magnetic Fields}

In this section we compile the minimum requirements for observing
quiet Sun IN fields with ATST and EST. With their 4m mirrors, these
telescopes will collect 64 times more photons than 50-cm telescopes
(if we assume the same photon efficiency), which will alleviate most
of the current limitations encountered in the study of quiet Sun IN
fields.

\begin{itemize}
\item[-] \emph{Spatial resolution}: On average, we know that the fill 
fraction is about 20\% for 120~km (0\farcs16) pixel size (e.g.,
\citealt{2007ApJ...670L..61O}). Thus, we need to increase the spatial
resolution up to 30 km (the diffraction limit of ATST and EST at
500~nm) to spatially resolve them.

\item[-] \emph{Temporal resolution}: a minimum time cadence may be 
given by the rms velocity of the magnetic elements and the spatial
resolution. Assuming a rms velocity of 1.5~km~s$^{-1}$ (measured
with a 1-meter class telescope) and a spatial resolution of 30~km, the
magnetic elements would take about 20 seconds to move from one pixel
to the next. We shall set half of this time as the minimum cadence for
critical sampling of the motions in the case of diffraction-limited
observations. Note that since the measured rms velocity increases with
spatial resolution, a cadence of 10 seconds may be severely
underestimated. In addition, the rms velocity of the magnetic elements
also poses limitations for the exposure times that should be much
smaller than 10 seconds to prevent image degradation. If we take the
sound speed in the photosphere (about 8~km~s$^{-1}$) as a reference,
then all measurements have to be performed about five times faster,
i.e., in about two seconds. The high cadence will allow the study of
fast physical phenomena like cancellation events.

\item[-] \emph{Polarization sensitivity}: The number of photons collected 
per resolution element is independent of the telescope diameter at the
diffraction limit (see e.g., \cite{2003SPIE.4843..100K}). For this reason, 4-meter telescopes will have the same
polarization sensitivity as 50-centimeter telescopes, provided the
photon efficiency is maintained. Thus, taking the exposure times of
the Hinode spectropolarimeter as a reference, at the diffraction limit
ATST and EST will reach a s/n ratio of $\sim$~1000 with integration
times of five seconds. For the quiet Sun internetwork we need to go
beyond noise levels of $10^{-3} I_{\rm c}$. Thus, there are two
options: increase the exposure time or downgrade the spatial
resolution. For instance, we can reach $2.5
\times 10^{-4} \, I_{\rm c}$ with five second exposure times and
0\farcs1 resolution. Note that since it is expected that the
polarization signals will be larger because of the increased
resolution, it may be possible to fully characterize the IN fields
with s/n~$\sim$~1000. Thus, a better option may be to achieve a 
s/n of 1000 in less than a few tenths of a second at 0\farcs1 spatial
resolution.

\item[-] \emph{Field of view}: Which FOV is most appropriate for the study 
of quiet Sun magnetic fields? To investigate the dynamics and
interactions between magnetic elements, a FOV covering a minimum area
of 4\arcsec seems necessary. However, we have seen that there exist
clear connections between the dynamics and the spatial location within
supergranular cells. Therefore, the observation of a complete
supergranule (about 30\arcsec\/ diameter) would be highly desirable.
\end{itemize}

Is it possible for ATST and EST to meet all the above requirements?
With slit-based instruments only, the answer is negative. The reason
is that the spatial resolution of the observations should be kept
close to the diffraction limit of ATST and EST to advance our
understanding of IN magnetic fields. Hinode observations have shown
that a spatial resolution of 0\farcs3 is not enough to characterize
the magnetic processes taking place in the IN. But, at the diffraction
limit of a 4-meter telescope ($\sim$~0\farcs05), it takes more than
six minutes for a spectrograph to scan a 4\arcsec-wide area with a
s/n~$\sim$~1000. Six minutes is far above the minimum requirements set
by the current observations. One can make a compromise and reduce the
spatial resolution to about 0\farcs1, but it would still take about
1.5 minutes to scan the same area maintaining the s/n.

Therefore, systems capable of scanning a given FOV while maintaining
the polarimetric sensitivity and temporal resolution seem necessary if
one wants to fulfill the minimum requirements to characterize
internetwork fields, study their dynamics, and analyze the physical
processes in which they participate. These are, for instance,
multi-slit configurations, image slicing instruments, or fiber optics
arrays (see e.g., \citealt{2010AN....331..581S}). In particular, the
latter concept has been successfully implemented in the
SpectroPolarimetric Imager for the Energetic Sun (SPIES), an
instrument which observes 2D FOVs without compromising the spatial,
temporal, or spectral resolution (see Lin, this volume).

Another option is to use tunable filters instead of
spectrographs. However, although filtergraph observations can achieve
noise levels close to 10$^{-3}$ (e.g., \citealt{2011SoPh..268...57M})
and the determination of the magnetic field vector is possible from
such observations (e.g., \citealt{2010A&A...522A.101O}), the
uncertainties of the physical quantities inferred from these
instruments are larger than those derived from spectropolarimetric
measurements. Among other reasons, this is due to the fact that the
shapes of the spectral lines are strongly distorted by the passband of
filter-based polarimeters and the long times needed to scan the
spectral line.

\acknowledgements D.O.S. thanks the Japan Society for the Promotion of Science
(JSPS) for financial support through the postdoctoral fellowship
program for foreign researchers. Hinode is a Japanese mission developed and launched by ISAS/JAXA, with NAOJ as domestic partner and NASA and STFC (UK) as international partners. It is operated by these agencies in co-operation with ESA and NSC (Norway).

\bibliographystyle{asp2010}
\bibliography{orozcosuarez}

\end{document}